\documentclass[prb,amsmath,superscriptaddress,citeautoscript,twocolumn,
showkeys,showpacs,floatfix]{revtex4}

\usepackage{bm}
\usepackage{amssymb}
\usepackage{graphicx}

\begin{document}

\title{The noise spectra of a biased quantum dot}

\author{E. A. Rothstein}
\email{rotshtei@bgu.ac.il}

\affiliation{Department of Physics, Ben Gurion University, Beer
Sheva 84105, Israel}

\author{O. Entin-Wohlman}

\altaffiliation{Also at Tel Aviv University}

\affiliation{Department of Physics and the Ilse Katz Center for
Meso- and Nano-Scale Science and Technology, Ben Gurion University, Beer
Sheva 84105, Israel}

\affiliation{Albert Einstein Minerva Center for Theoretical
Physics, Weizmann Institute of Science, Rehovot 76100, Israel}

\author{A. Aharony}

\altaffiliation{Also at Tel Aviv University}

\affiliation{Department of Physics and the Ilse Katz Center for
Meso- and Nano-Scale Science and Technology, Ben Gurion University, Beer
Sheva 84105, Israel}

\date{\today}

\begin{abstract}
The noise spectra associated with correlations of the current
through a single level quantum dot, and with the charge
fluctuations on the dot, are calculated for a finite bias voltage.
The results turn out to be sensitive to the asymmetry of the dot's
coupling to the two leads. At zero temperature, both spectra
exhibit two or four steps (as a function of the frequency),
depending on whether the resonant level lies outside or within the
range between the chemical potentials on the two leads. In
addition, the low frequency shot-noise exhibits dips in the charge
noise and dips, peaks, and discontinuities in the derivative of
the current noise. In spite of some smearing, several of these
features persist at finite temperatures, where a dip can also turn
into a peak.

\end{abstract}


\maketitle

\section{Introduction}


Ten years ago, Landauer coined the phrase ``the noise is the
signal".\cite{LANDU}  Indeed, the noise spectrum of electronic
transport through mesoscopic systems provides invaluable
information on the physics which governs this
transport.\cite{book,BB} 
The
noise spectrum is given by the Fourier transform of the
current-current correlation.  The {\em un-symmetrized} noise spectrum is defined as
\cite{book}
\begin{align}
C^{}_{\alpha\alpha '}(\omega )=\int _{-\infty} ^{\infty} dt
e^{-i\omega t}\langle \delta\hat{I}^{}_{\alpha}(t)\delta
\hat{I}^{}_{\alpha '}(0)\rangle\ ,\label{CDEF}
\end{align}
where $\alpha$ and $\alpha '$ mark the leads, which carry the
current from  the electron reservoirs to the mesoscopic system. In
Eq. (\ref{CDEF}), $\delta\hat{I}^{}_\alpha\equiv
\hat{I}^{}_\alpha-\langle\hat{I}^{}_\alpha\rangle$, where
$\hat{I}^{}_\alpha$ is the current operator in lead $\alpha$, and
the average (denoted by $\langle\ldots\rangle$) is taken over the
states of the reservoirs (see below). At finite frequencies, this
quantity is very sensitive to the locations where those currents
are monitored. When $\alpha =\alpha '$, Eq. (\ref{CDEF}) gives the
{\em auto-correlation} function, while for $\alpha\neq\alpha '$ it
yields the {\em cross-correlation} one. Clearly, $C_{\alpha
\alpha'}(\omega) = C_{\alpha' \alpha} ^{\ast} (\omega)$, and
consequently the auto-correlation function is real. Some papers
prefer to analyze the {\it symmetrized} noise spectrum, defined as
$[C_{\alpha \alpha'}(\omega)+C_{\alpha' \alpha}(-\omega)]/2$.
However, as we discuss below, this spectrum may miss some
important features. Particular measurements require the
calculation of different combinations of the $C^{}_{\alpha\alpha
'}$'s.

In this article we calculate the various current correlations,
$C^{}_{\alpha\alpha '}(\omega )$, for the simplest mesoscopic
system, i.e.  a single level quantum dot connected to two electron
reservoirs via leads $L$ and $R$. The latter are kept at different
chemical potentials, $\mu _L$ and $\mu _R$. The potential
difference,
\begin{align}
V=(\mu ^{}_L - \mu^{} _R)/e\ ,\label{POT}
\end{align}
represents the bias voltage applied to the dot. It is convenient
to measure energies relative to the common Fermi energy, $(\mu
^{}_L + \mu^{} _R)/2$. Setting this energy to zero, we have
$\mu^{}_L=-\mu^{}_R=eV/2$. Having two leads, one can consider two
auto-correlation functions, $C^{}_{LL}(\omega)$ and
$C^{}_{RR}(\omega)$, and two cross-correlation functions,
$C^{}_{LR}(\omega)$ and $C^{}_{RL}(\omega)$.

The operator of the net
current going through the dot is given by
\begin{align}
\hat{I}=(\hat{I}^{}_L-\hat{I}^{}_R)/2\ .\label{Im}
\end{align}
With a finite bias voltage, $\langle\hat{I}\rangle$ is not
necessarily zero. The noise associated with $\hat{I}$ is then
given by
\begin{align}
&C^{(-)}(\omega )= \frac{1}{4}\Bigl (C^{}_{LL}(\omega
)+C^{}_{RR}(\omega )-C^{}_{LR}(\omega )-C^{}_{RL}(\omega )\Bigr )\
.\label{CII}
\end{align}
Since $C_{LR}^{}(\omega )=C^{\ast}_{RL}(\omega )$, $C^{(-)}(\omega
)$ is real. Alternatively, one could also consider the difference
between the currents flowing into the dot from the two leads,
\begin{align}
\Delta\hat{I}=(\hat{I}^{}_L+\hat{I}^{}_R)/2\ ,
\end{align}
for which $\langle \Delta\hat{I} \rangle=0$.\cite{BB} The
fluctuations in this difference account for the fluctuations in
the {\it net charge} accumulating on the dot. The noise associated
with this charge is given by
\begin{align} &C^{(+)}(\omega )= \frac{1}{4}\Bigl
(C^{}_{LL}(\omega )+C^{}_{RR}(\omega )+C^{}_{LR}(\omega
)+C^{}_{RL}(\omega )\Bigr )\ .\label{Cp}
\end{align}

Earlier theoretical papers considered various aspects of noise correlations in mesoscopic systems.
Some of these studies analyzed only the low frequency
limit of the spectrum, which reduces to the Johnson-Nyquist noise
at equilibrium (i.e. at zero bias voltage) and to the shot noise
at a finite bias. Specifically, Chen and Ting\cite{CHEN} studied
the un-symmetrized noise associated with the net terminal current
[our Eq. (\ref{Im})], and found a Lorentzian peak around zero
frequency for a bias which is larger than the resonance level width. Averin\cite{AVERIN} then studied the
shot noise for any value of the bias, but considered only the
symmetrized noise at the zero frequency limit. Engel and
Loss\cite{ENGEL} extended these results to finite frequencies and
to the un-symmetrized noise, but considered only the
auto-correlation function. They found steps at particular
frequencies. For the two level dot they also found a dip in the
auto-correlation noise around zero frequency, which they
attributed to ``the charging effect of the dot". 




Recently, two of us participated in a detailed analysis of
$C^{(-)}(\omega)$ in the limit of zero temperature and zero
bias.\cite{EWIGA} Ignoring interactions and capacitance effects,
which might add correlations among the currents at relatively high
frequencies,\cite{BB,BUT1,BUT2} it is convenient to use the single
electron  scattering formalism for obtaining explicit expressions
for the noise spectrum.\cite{but0,lev} Similar to Ref.
~\onlinecite{ENGEL}, Ref. ~\onlinecite{EWIGA} found that the
current noise spectrum $C^{(-)}(\omega)$ has a step structure as a
function of the frequency, with the step edges located roughly at
energies corresponding to the resonances of the quantum dot. It
was consequently suggested that the noise spectrum can be used to
probe the resonance levels of the dot. For the two level quantum
dot, Ref. ~\onlinecite{EWIGA} also found dips in the noise
spectrum, which appeared when the Fermi energy was between the two
levels and deepened upon increasing the asymmetry of the coupling
of the dot to the leads. 

The present paper generalizes Ref. ~\onlinecite{EWIGA}, by
introducing
 a finite bias and and a finite temperature, and by considering also the charge noise. In particular, we discuss the interesting dependence
of the various spectra on the spatial asymmetry, denoted by
\begin{align}
a=\frac{\Gamma_{R}^{}-\Gamma_{L}^{}}{\Gamma_{R}^{}+\Gamma_{L}^{}}\
,\ \ -1\leq a\leq 1\ , \label{a}
\end{align}
where $\Gamma_L$ ($\Gamma_R$) denote the broadening of the
resonance on the dot due to its coupling with the left (right)
lead. 
 We find that the single
step appearing in $C^{(\pm)}(\omega )$ in the absence of the bias
\cite{EWIGA}  splits in the presence of $V$  into two steps when
the resonance energy $\epsilon^{}_d$ is not between the two
chemical potentials ($|\epsilon^{}_d|>|eV/2|$) and into four steps
when it is
within that range. 
In addition, $C^{(+)}(\omega)$ vanishes at $\omega=0$, exhibiting
a dip in the shot noise around $\omega=0$. For
$|\epsilon^{}_d|<|eV/2|$ and for $0<|a|<1$ there also appears a
{\em peak} in $C^{(-)}(\omega)$ at $\omega=0$. At zero temperature
and close to $|eV/2|=|\epsilon^{}_d|$ we also find a discontinuity
in the slope of $C^{(-)}(\omega)$. Many of these features are
smeared as the temperature $T$ increases. However, both
$C^{(+)}(\omega)$ and $C^{(-)}(\omega)$ still exhibit dips and
peaks near $\omega=0$ even at $T>0$.



Our paper is divided into two main sections. In Sec. II we discuss
some general properties of the noise spectra, present a short
review of the scattering matrix formalism, and from that derive
the various noise spectra for a single level quantum dot. In the
following section we analyze both $C^{(+)}(\omega)$ and
$C^{(-)}(\omega)$, with and without a bias voltage and at both
zero and non-zero temperatures.  The last section summarizes
our results. 

\section{The noise spectra}

\subsection{General relations}

We begin our discussion by describing several general properties
of the noise spectrum (which also hold  for interacting systems).
The physical meaning of the auto-correlation function is revealed
upon  re-writing  it in the form \cite{GAVISH2}
\begin{align}
C^{}_{\alpha \alpha }(\omega )=2\pi\sum_{i,f}P_{i}| \langle
i|\delta\hat{I}^{}_{\alpha}|f\rangle |^{2}\delta
(E^{}_{i}-E^{}_{f}-\omega )\ .\label{AUTO}
\end{align}
Here, $|i\rangle$ and $|f\rangle$ are the initial and final states
of the whole system (the dot and its leads), with the
corresponding energies $E_{i}$ and $E_{f}$. In Eq. (\ref{AUTO}),
$P_i$ is the probability for the system to be in the initial state
$|i\rangle$.  It is now seen that  the auto-correlation is the
rate (as given by the Fermi golden-rule) by which the system
absorbs energy from a monochromatic electromagnetic field of
frequency $\omega$. The symmetrized noise spectrum mixes
absorption and emission, and thus loses the separation between the
two.

At zero frequency, $\omega =0$, the auto-correlation and the
cross-correlation are related to one another. This follows from
charge conservation.\cite{AVI} The equation of motion for
$\delta\hat{ n}(t)$, the fluctuation of the occupation operator on
the dot, is given by
\begin{align}
\label{OCCUPATION} e \frac{d \delta \hat{n}(t)}{dt} = \delta
\hat{I}^{}_{L}(t) + \delta \hat{I}^{}_{R}(t) \ .
 \end{align}
[Note that in our convention,  the currents flowing in the left
(right) lead, $\hat{I}^{}_{L}$ ($\hat{I}^{}_{R}$), are directed
{\it towards} the dot.] Equation (\ref{OCCUPATION}) implies that
\begin{align}
&C^{}_{L\alpha}(0)+C^{}_{R\alpha }(0)=e\int_{-\infty}^{\infty}dt \langle
\frac{d \delta \hat{n}(t)}{dt} \delta\hat{I}^{}_{\alpha }(0)\rangle\nonumber\\
&=e\lim_{\tau\rightarrow\infty}\langle\delta\hat{n}(\tau )
\delta\hat{I}^{}_{\alpha}(0)-
\delta\hat{n}(-\tau )\delta\hat{I}^{}_{\alpha}(0)\rangle\ .
\end{align}
At steady-state, assuming no long-term memory, we have
$\lim_{\tau\rightarrow\pm\infty}\langle\delta\hat{n}(\tau )
\delta\hat{I}^{}_{\alpha}(0)\rangle=\langle\delta\hat{n}\rangle\langle\hat{I}^{}_{\alpha}\rangle$,
and therefore $C^{}_{L\alpha}(0)+C^{}_{R\alpha }(0)=0$.\cite{AVI}
As a result,
\begin{align}
C^{}_{LL}(0)=-C^{}_{RL}(0)\ ,\ \ C^{}_{RR}(0)=-C^{}_{LR}(0)\ .
\end{align}
Moreover, since $C^{}_{LL}(0)$ and $C^{}_{RR}(0)$ are real and
positive [see Eq. (\ref{AUTO})], it follows that the
zero-frequency cross-correlations are real as well, but negative.
Since the cross-correlations are real, one has
$C^{}_{LR}(0)=C^{}_{RL}(0)$, and therefore
\begin{align}
C^{}_{LL}(0)=C^{}_{RR}(0)=-C^{}_{LR}(0)=-C^{}_{RL}(0)\ .\label{JN}
\end{align}
In particular, this implies that $C^{(+)}(0)=0$ and
$C^{(-)}(0)=C^{}_{LL}(0)$. At zero bias, those  are just the
Nyquist-Johnson relations, $C^{}_{LL}(0)=C^{}_{RR}(0)
=-C^{}_{LR}(0)=-C^{}_{RL}(0)=k_{\rm B}TG(0)$, where $G(0)$ is the
dc conductance of the dot.

\subsection{The noise spectrum in the scattering formalism}

When electron-electron interactions are ignored,
one may use the (single-particle) scattering matrix of the dot
to obtain an expression for
the noise spectrum in terms of the scattering
matrix elements. This has been accomplished in Refs. ~ \onlinecite{BUT1} and ~ \onlinecite{BUT2}.

In the scattering formalism, one expresses the current operator
$\hat{I}$ in terms of creation [$\hat{a}^{\dagger}_{\alpha}(E)$]
and annihilation [$\hat{a}^{}_{\alpha}(E)$] operators of the
electrons in the reservoir connected to terminal $\alpha$. These
operators are normalized such that
\begin{align}
\langle \hat{a}^{\dagger}_{\alpha}(E)\hat{a}^{}_{\alpha
'}(E')\rangle =\delta^{}_{\alpha\alpha '}\delta (E-E')f^{}_{\alpha
}(E)\ ,\label{AVEA}
\end{align}
where $f_{\alpha}(E)\equiv [\exp (E-\mu_{\alpha} )/k_{\rm
B}T+1]^{-1}$ is the Fermi distribution in reservoir $\alpha$ which
is held at the chemical potential $\mu_{\alpha}$. The explicit
form for the current operator (using units in which $\hbar=1$) is
\cite{but0}
\begin{align}
\hat{I}_{\alpha}^{}(t)&=\frac{e}{2\pi}\int _{-\infty}
^{\infty} dE\int _{-\infty} ^{\infty} dE'
e^{i(E-E')t}\nonumber\\
&\times \sum_{\gamma\gamma '}A^{}_{\gamma \gamma '}(\alpha ,E,
E')\hat{a}^{\dagger}_{\gamma}(E)\hat{a}^{}_{\gamma '}(E')\ ,
\label{IKO}
\end{align}
with
\begin{align}
A^{}_{\gamma\gamma '}(\alpha , E, E') =\delta^{}_{\gamma\gamma
'}\delta^{}_{\alpha \gamma}-S^{\ast}_{\alpha\gamma}(E)S^{}_{\alpha
\gamma '}(E')\ ,\label{AA}
\end{align}
where Greek letters denote the lead indices and
$S^{}_{\alpha\gamma}$ are the elements of the scattering
matrix characterizing the dot.

Inserting the expression for the current operator, Eq.
(\ref{IKO}), into Eq. (\ref{CDEF}) and calculating the averages
according to Eq. (\ref{AVEA}), we find \begin{align} \label{AA'}
C_{\alpha \alpha '}(\omega) &= \frac{e^2}{2 \pi} \int _{-\infty}
^{\infty} dE  \sum _{\gamma \gamma '}
F^{\alpha\alpha'}_{\gamma\gamma'}(E,\omega)\nonumber\\
&\times f_{\gamma} (E+\omega ) (1 - f_{\gamma '} (E))\ ,
\end{align}
where \begin{align}
F^{\alpha\alpha'}_{\gamma\gamma'}&(E,\omega)\equiv A _{\gamma
\gamma '} (\alpha,E+\omega ,E) A_{\gamma'
\gamma} (\alpha', E, E+\omega)\nonumber\\
&=A _{\gamma \gamma '} (\alpha,E+\omega ,E) A^\ast_{\gamma
\gamma'} (\alpha', E+\omega,E)\ .
\end{align}
It is straightforward to verify, using the unitarity of the
scattering matrix, that the zero-frequency relations (\ref{JN})
are obeyed by the form (\ref{AA'}).  Another limit of Eq.
(\ref{AA'}) is obtained upon neglecting the energy dependence of
the scattering matrix elements. Then (at zero temperature and for
$\omega > 0$) one retrieves the well-known result \cite{book}
$C_{LL}(\omega )=(e^{2}/2\pi ){\cal T}(1-{\cal T})(eV-\omega
)\Theta (eV-\omega )$, where ${\cal T}$ is the transmission of the
dot.

We next discuss the correlation functions $C^{(\pm)}(\omega)$,
Eqs. (\ref{CII}) and (\ref{Cp}). Upon inserting Eqs. (\ref{AA})
and (\ref{AA'}) into Eqs. (\ref{CII}) and (\ref{Cp}) we obtain
\begin{align}
\label{NOISE} C^{(\pm)}(\omega
)&=\frac{e^{2}}{8\pi}\int_{-\infty}^{\infty}dE \sum_{\gamma\gamma
'}F^{(\pm)}_{\gamma\gamma '}
(E,\omega )\nonumber\\
&\times f^{}_{\gamma}(E+\omega )(1-f^{}_{\gamma '}(E)),
\end{align}
where
\begin{align}
\label{FAA'} &F^{(\pm)}_{LL}(E,\omega )=\Big
|1-S^{\ast}_{LL}(E+\omega )S^{}_{LL}(E)
\nonumber\\
&\ \ \ \ \ \ \ \ \ \mp S^{\ast}_{RL}(E+\omega )S^{}_{RL}(E)\Big |^{2}\ ,\nonumber\\
&F^{(\pm)}_{LR}(E,\omega)=\Big |S^{\ast}_{LL}(E+\omega )S^{}_{LR}(E) \nonumber\\
&\ \ \ \ \ \ \ \ \ \pm S^{\ast}_{RL}(E+\omega )S^{}_{RR}(E)\Big
|^{2}\ .
\end{align}
The other correlations,  $F^{(\pm)}_{RR}(E,\omega )$ and
$F^{(\pm)}_{RL}(E,\omega )$, are obtained from these expressions
upon interchanging $L\leftrightarrow R$. In this way we divide the
correlation functions $C^{(\pm)}(\omega)$ according to the
separate contributions of the various processes:
$F^{(\pm)}_{LL}(E,\omega )$ and $F^{(\pm)}_{RR}(E,\omega )$
describe {\em intra-lead} transitions of the electron, while
$F^{(\pm)}_{LR}(E,\omega )$ and $F^{(\pm)}_{RL}(E,\omega )$ give
the contributions of the {\em inter-lead} processes. The actual
contribution of each process to $C^{(\pm)}(\omega )$ is determined
by the relevant product of the Fermi functions. In particular, at
zero temperature ($T=0$), this product vanishes everywhere except
on a finite segment of the energy axis. Finite temperatures
broaden and smear the limits of this section, while the
application of the bias voltage may shift it along the energy axis
or change its length. 

\subsection{A single level dot}

In our simple configuration, the dot is represented by a single
energy level denoted $\epsilon^{}_{d}$.  As mentioned, we denote
the broadening due to the coupling with the left lead by
$\Gamma_{L}$, and that due to the coupling with the right one by
$\Gamma_{R}$, such that the total width of the energy level on the
dot is
\begin{align}
\Gamma =\Gamma^{}_{L}+\Gamma^{}_{R}\ .
\end{align}
In this model the scattering matrix takes the form
\begin{align}
\label{SCATTERING1}
S(E)&=\left[
\begin{array}{cc}
S^{}_{LL}(E) & S^{}_{LR}(E) \\
S^{}_{RL}(E) & S^{}_{RR}(E)
\end{array}\right ]\nonumber\\
&=-1+ig(E)\left [\begin{array}{cc}\Gamma^{}_{L}&\sqrt{\Gamma_{L}^{}\Gamma_{R}^{}}\\
\sqrt{\Gamma_{L}^{}\Gamma_{R}^{}}&\Gamma_{R}^{}\end{array}\right ]\ ,
\end{align}
where
$g(E)$ is the Breit-Wigner resonance formed by the dot,
\begin{align}
g(E) = \frac{1}{E - \epsilon _d + i \Gamma /2}\ .\label{G}
\end{align}
(Assuming the scattering to take place at about the Fermi energy,
we have discarded the energy dependence of  the resonance partial
widths.)

Since the dot forms a Breit-Wigner resonance, it is useful to
express the functions $F^{(\pm)}_{\alpha \alpha '}(E,\omega )$,
Eq. (\ref{FAA'}), in terms of the resonance phase $\delta (E)$,
defined by \cite{NG}
\begin{align}
{\rm cot}\delta (E)=\frac{2}{\Gamma}(\epsilon^{}_{d}-E)\ ,\label{DELDEF}
\end{align}
such that $g(E)$, Eq. (\ref{G}), becomes
\begin{align}
g(E)=-i\frac{2}{\Gamma}\sin\delta (E) e^{-i\delta (E)}\ .
\end{align}
Clearly, $|g(E)|^{2}$ is peaked around $E=\epsilon^{}_{d}$,  and
the phase $\delta (E)$ changes from 0 to $\pi$ within a range of
width $\Gamma$ around this resonance.

Using the identities
\begin{align}
g(E)\pm g^\ast(E+\omega)=a^{}_\pm g(E)g^\ast(E+\omega),
\end{align}
where
\begin{align}
a^{}_-&=\omega-i\Gamma=\frac{\Gamma\sin[\delta(E+\omega)-\delta(E)]}
{2\sin\delta(E)\sin\delta(E+\omega)}-i\Gamma,\nonumber\\
a^{}_+&=2(E-\epsilon^{}_d)+\omega=-\frac{\Gamma}{2}[\cot\delta(E)+\cot\delta(E+\omega)]\nonumber\\
&=-\frac{\Gamma\sin[\delta(E+\omega)+\delta(E)]}
{2\sin\delta(E)\sin\delta(E+\omega)}\ , \end{align} we find
\begin{align}
F^{(+)}_{\gamma\gamma'}(E,\omega)&=\Gamma_\gamma\Gamma_{\gamma'}|g(E)g(E+\omega)|^2\omega^2\nonumber\\
&=\frac{16\Gamma_\gamma\Gamma_{\gamma'}}{\Gamma^4}\sin^2\delta(E)\sin^2\delta(E+\omega)\omega^2\nonumber\\
&=\frac{4\Gamma_\gamma\Gamma_{\gamma'}}{\Gamma^2}\sin^2[\delta(E+\omega)-\delta(E)]\
\label{Fpp}
\end{align}
and
\begin{widetext}
\begin{align}
F^{(-)}_{LR}(E,\omega
)&=F^{(+)}_{LR}+\Gamma_L\Gamma_R[(\Gamma_L-\Gamma_R)^2+4(E-\epsilon^{}_d)(E-\epsilon^{}_d+\omega)]|g(E)g(E+\omega)|^2\nonumber\\
&=\frac{16\Gamma^{}_{L}\Gamma^{}_{R}}{\Gamma^{4}} \Bigl
((\Gamma^{}_{L}-\Gamma_{R}^{})^{2}\sin^{2}\delta (E)\sin^{2}\delta
(E+\omega ) +\frac{\Gamma^{2}}{4}\sin^{2}\bigl [\delta (E+\omega
)+\delta (E)\bigr ]\Bigr )\nonumber\\
F^{(-)}_{LL}(E,\omega
)&=F^{(+)}_{LL}+4\Gamma_L^2\Gamma_R^2|g(E)g(E+\omega)|^2=\frac{16\Gamma_L^2}{\Gamma^4}(4\Gamma_R^2+\omega^2)\sin^{2}\delta
(E)\sin^{2}\delta (E+\omega )\ .\label{integrands}
\end{align}
\end{widetext}
Hence, each of the integrands appearing in  Eq.  (\ref{FAA'})
includes two resonances, around $E\simeq \epsilon^{}_{d}$, and
around $E+\omega \simeq\epsilon^{}_{d}$. These resonances
determine the dependence of the noise spectrum on the frequency.
In the next section we study this dependence, allowing for a
possible asymmetry between the left and right couplings, Eq.
(\ref{a}).

For maximal asymmetry, $|a|=1$, i.e. when one of the two leads is
decoupled from the dot, we have $\Gamma_L\Gamma_R=0$, and
therefore the inter-lead correlations vanish. In this case we have
$C^{(+)}(\omega)=C^{(-)}(\omega)$.

Another general feature of Eq. (\ref{NOISE}) is that
$C^{(\pm)}(\omega)$ is invariant under the simultaneous sign
change of $V$ and of the asymmetry $a$. Therefore, we present
below only results for $V>0$. In addition, the noise is also
symmetric under the simultaneous sign change of $\epsilon^{}_d$
and of the asymmetry parameter $a$, and therefore we present
results only for $\epsilon^{}_d<0$, i.e. when the localized level
on the dot
is placed below the common Fermi energy of the reservoirs.

As seen from Eq. (\ref{Fpp}), the functional form of the charge
noise $C^{(+)}(\omega)$ is much simpler than those for the other
spectra. Therefore, we start our presentations below with a
discussion of $C^{(+)}(\omega)$. It also turns out to be useful to
discuss the cross-correlation noise,
\begin{align}
C^{(\times)}(\omega)\equiv
C^{(+)}(\omega)-C^{(-)}(\omega)=[C_{LR}(\omega)+C_{RL}(\omega)]/2.\label{eq31}
\end{align}
As we shall see below, $C^{(\times)}(\omega)$ is usually small,
and it has interesting structure only in the shot-noise regime
near $\omega=0$, where we find differences between
$C^{(+)}(\omega)$ and $C^{(-)}(\omega)$.

\section{Results}

\label{RES}

\subsection{The unbiased dot}

\label{UNBIASED}

The unbiased noise $C^{(-)}(\omega)$ has been treated in Ref.
~\onlinecite{EWIGA} for $T=0$.  Here we extend these results also
to $C^{(+)}(\omega)$ and to finite $T$.  When the potential
(\ref{POT}) vanishes, the two Fermi distributions become
identical, and Eq. (\ref{NOISE}) becomes
\begin{align}
\frac{8\pi}{e^{2}}C^{(\pm)}(\omega )&=\int^{\infty }_{-\infty}dE
f(E+\omega)[1-f(E)]\nonumber\\
&\times \sum_{\gamma\gamma'}F^{(\pm)}_{\gamma\gamma'}(E,\omega)\ .
\end{align}
In particular, Eq. (\ref{Fpp}) now implies that
\begin{align}
&\frac{2\pi}{e^{2}}C^{(+)}(\omega )=\int^{\infty }_{-\infty}dE
f(E+\omega)[1-f(E)]\nonumber\\
&\times \sin^2[\delta(E)-\delta(E+\omega)]\ ,
\end{align}
independent of the asymmetry $a$.

\begin{figure}[h]
\includegraphics[width=2.5in]{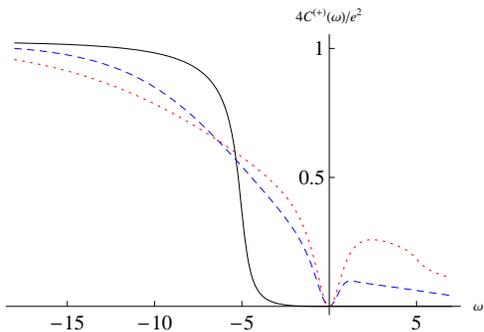}\\
\caption{(Color online) The noise spectrum $C^{(+)}(\omega)$ for
$V=0$. Here $\epsilon^{}_{d}=-5$ (energies and noise are measured
in units of $\Gamma$). The three curves correspond to $k_BT=0$
(black continuous line), $k_BT=3$ (blue dashed line) and $k_BT=5$
(red dotted line).} \label{Cplus}
\end{figure}

Consider first $T=0$. In this case, the integration is over
$0<E<-\omega$, and all the noise functions vanish for $\omega>0$.
The phase $\delta(E)$ [Eq. (\ref{DELDEF})] increases abruptly from
0 to $\pi$ as $E$ crosses the resonance at $E\sim \epsilon^{}_d$.
For $\epsilon^{}_d<0$, this resonance is out of the integration
range, so that $\delta(E)$ does not vary much within this range.
Similarly, $\delta(E+\omega)$ changes abruptly near $E\sim
\epsilon^{}_d-\omega$. This resonance enters the range of
integration when $\omega$ goes below $\epsilon^{}_d$. Using the
relation $2\sin^2\delta dE=\Gamma d\delta$, we conclude that the
integral over the resonance yields $\Gamma \pi/2$, ending up with
a step $\Gamma$ in $4C^{(+)}/e^2$. This step agrees with the
calculations of $C^{}_{LL}$ in Ref. ~\onlinecite{ENGEL}. In fact,
the variation of $4C^{(+)}(\omega)/e^2$ follows that of
$\Gamma\delta(\omega)/\pi$. \cite{EWIGA} Indeed, this step is
exhibited by the full calculation of the integral, shown by the
full line in Fig. \ref{Cplus}. A similar argument applies when
$\epsilon^{}_d>0$, when the step arises due to the resonance at
$E\sim\epsilon^{}_d$. Finite temperature smears the boundaries of
the integration, and thus smears the step, extending its tail to
 $\omega>0$, see Fig. \ref{Cplus}. However, as stated
following Eq. (\ref{JN}), we must have $C^{(+)}(0)=0$. For small
$|\omega|$, the second line in Eq. (\ref{Fpp}) implies that
\begin{align} &\frac{2\pi}{e^{2}}C^{(+)}(\omega )\approx \frac{4\omega^2}{\Gamma^2} \int^{\infty
}_{-\infty}dE f(E)[1-f(E)] \sin^4\delta(E)\
.
\end{align}
Thus, $C^{(+)}(\omega)$ has a parabolic-like dip around
$\omega=0$, as can indeed be seen in Fig. \ref{Cplus}. At low
temperatures we can replace $f(E)[1-f(E)]/(k_BT)$ by the Dirac
delta function, and then we find $(4\pi/e^2)C^{(+)}(\omega)\approx
8(\omega^2/\Gamma)^2 k_BT\sin^4\delta(0)$. Thus, the parabola
broadens with decreasing $T$, and vanishes at $T=0$.

\begin{figure}[h]
\includegraphics[width=2.5in]{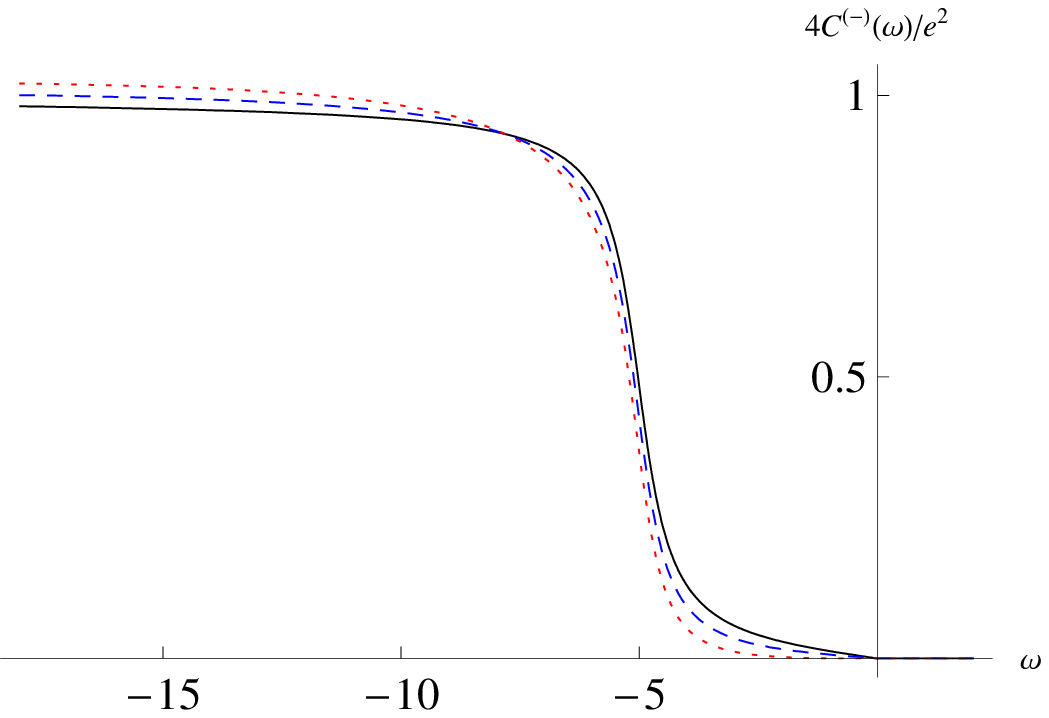}\\
\includegraphics[width=2.5in]{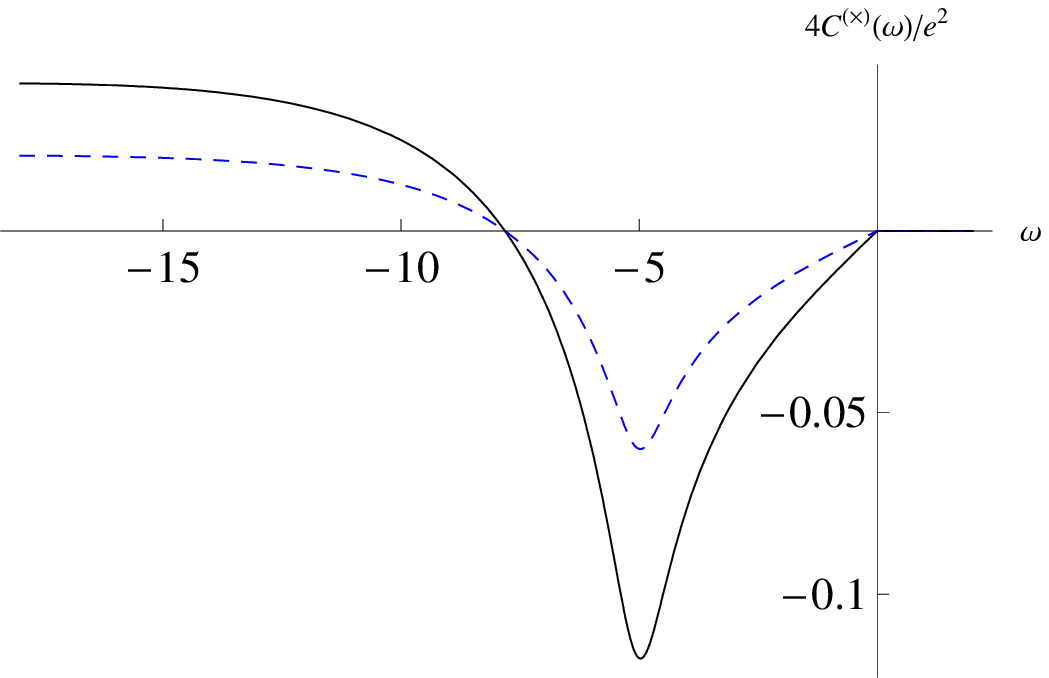}\\
\caption{(Color online) Upper panel: the noise spectrum
$C^{(-)}(\omega)$ for $V=0$ and $T=0$. Here $\epsilon^{}_{d}=-5$
(energies and noise are measured in units of $\Gamma$). The three
curves correspond to zero asymmetry $a=0$ (the black continuous
curve), $a= 0.7$ (the blue dashed curve), $a= 1$ (the red dotted
curve). Results are independent of the sign of $a$. Lower panel:
the cross-correlation function $C^{(\times)}(\omega)$ for $a=0$
(black continuous curve) and $a=0.7$ (blue dashed curve).}
\label{NOBIAS}
\end{figure}

\begin{figure}[h]
\vspace{5mm}
\includegraphics[width=2.5in]{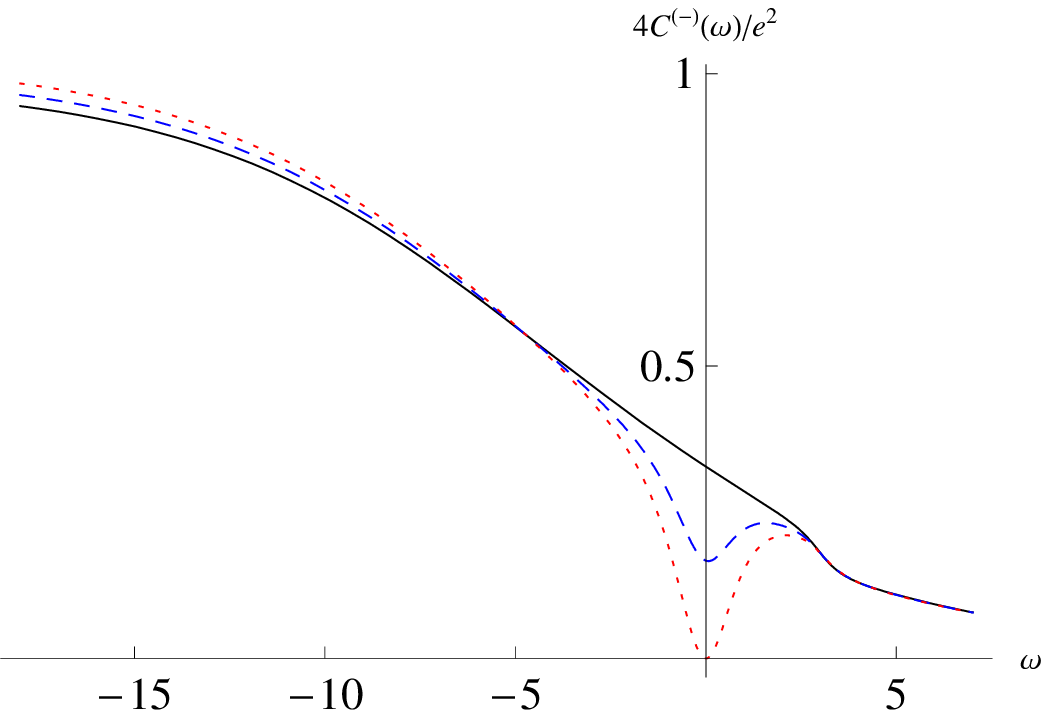}\\
\includegraphics[width=2.5in]{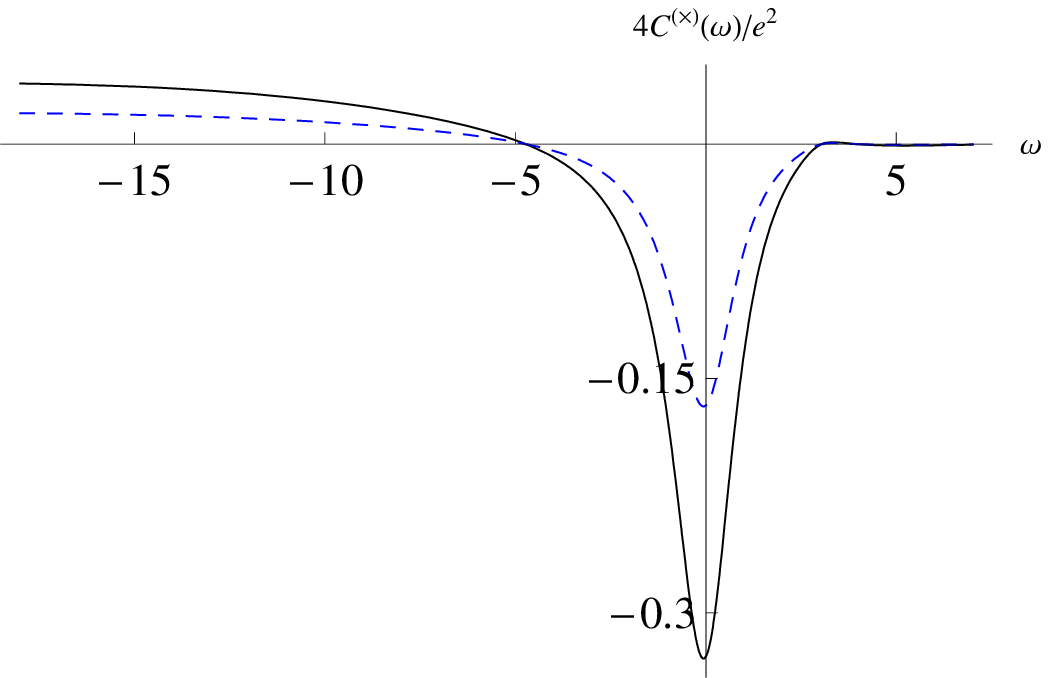}\\
\caption{(Color online) Same as Fig. \ref{NOBIAS}, but for
$k_BT=4$. } \label{finT}
\end{figure}

At maximal asymmetry, $|a|=1$, we saw that
$C^{(-)}(\omega)=C^{(+)}(\omega)$. As seen from Eqs.
(\ref{integrands}) and (\ref{eq31}), the difference
$C^{(\times)}(\omega)$ involves
$\Gamma_L^{}\Gamma_R^{}=\Gamma^2(1-a^2)/4$, and therefore it does
not depend on the sign of $a$, and it increases as $|a|$ decreases
from $|a|=1$ to $a=0$. This difference involves an integration
over the product $|g(E)g(E+\omega)|^2$, which is small everywhere,
unless the two resonances overlap. Therefore, the
cross-correlation $C^{(\times)}(\omega)$ can be relatively large
only for $|\omega|\leq 2\Gamma$.
 Figure \ref{NOBIAS} shows $C^{(-)}(\omega)$ and
$C^{(\times)}(\omega)$ for $V=0$, $T=0$ and several values of
$|a|$. Indeed, $C^{(\times)}(\omega)$  has a small negative peak
at $\omega \simeq \epsilon^{}_d$, with the largest magnitude for
$a=0$. The cross-correlation function $C^{(\times)}(\omega)$ then
vanishes at some negative frequency, and reaches a small positive
plateau for large negative $\omega$. Figure \ref{finT} shows the
same results at $k_BT=4$. Interestingly, at this temperature the
negative dip in $C^{(\times)}(\omega)$ moved to the vicinity of
$\omega=0$. As a result, $C^{(-)}(\omega)$ exhibits a dip around
$\omega=0$, whose depth decreases with decreasing $|a|$ until it
disappears for $a=0$.

\subsection{The biased dot at $T=0$}

\label{BIASED}

\begin{widetext}

\begin{figure}[h]
\includegraphics[width=2.2in]{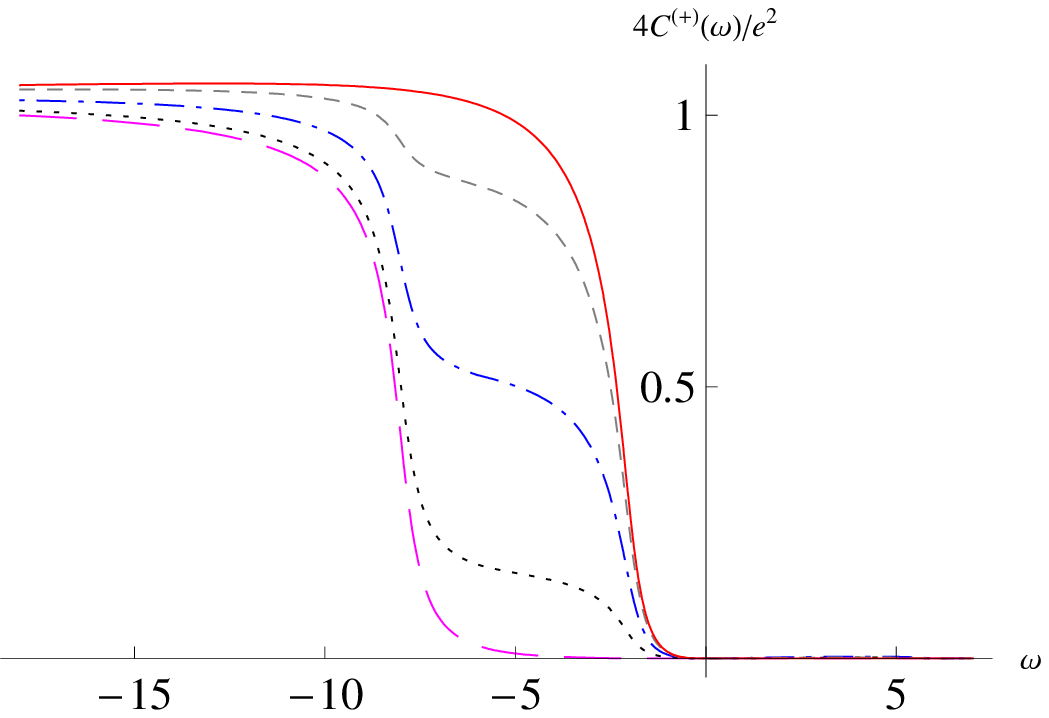}
\hspace{0.5cm}\includegraphics[width=2.2in]{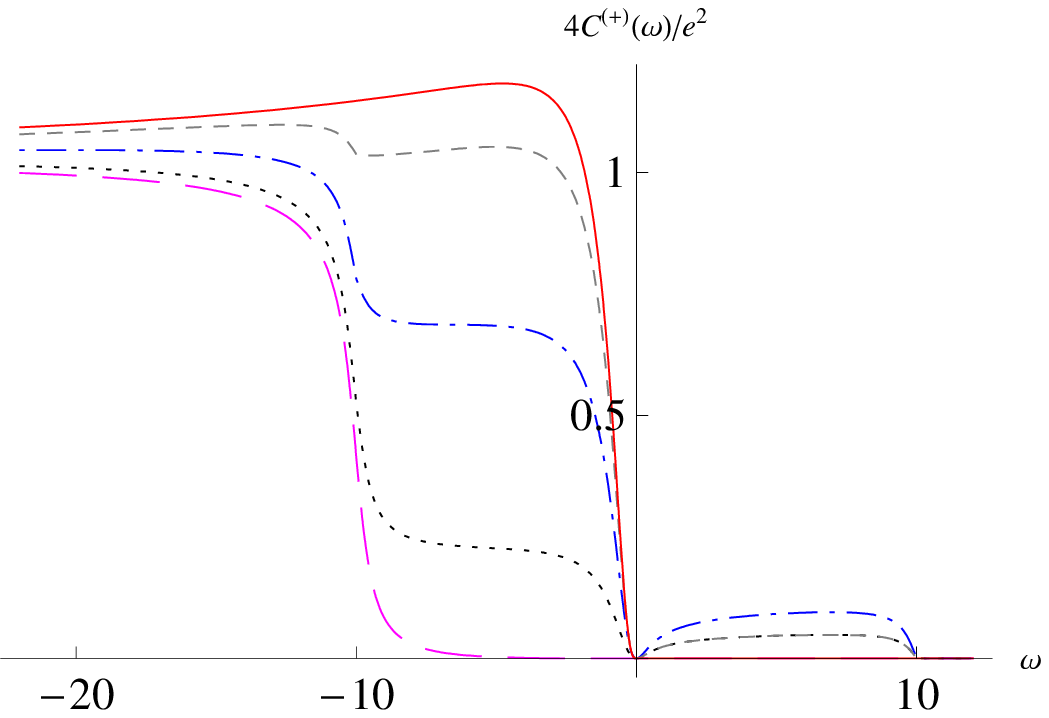}
\hspace{0.5cm}\includegraphics[width=2.2in]{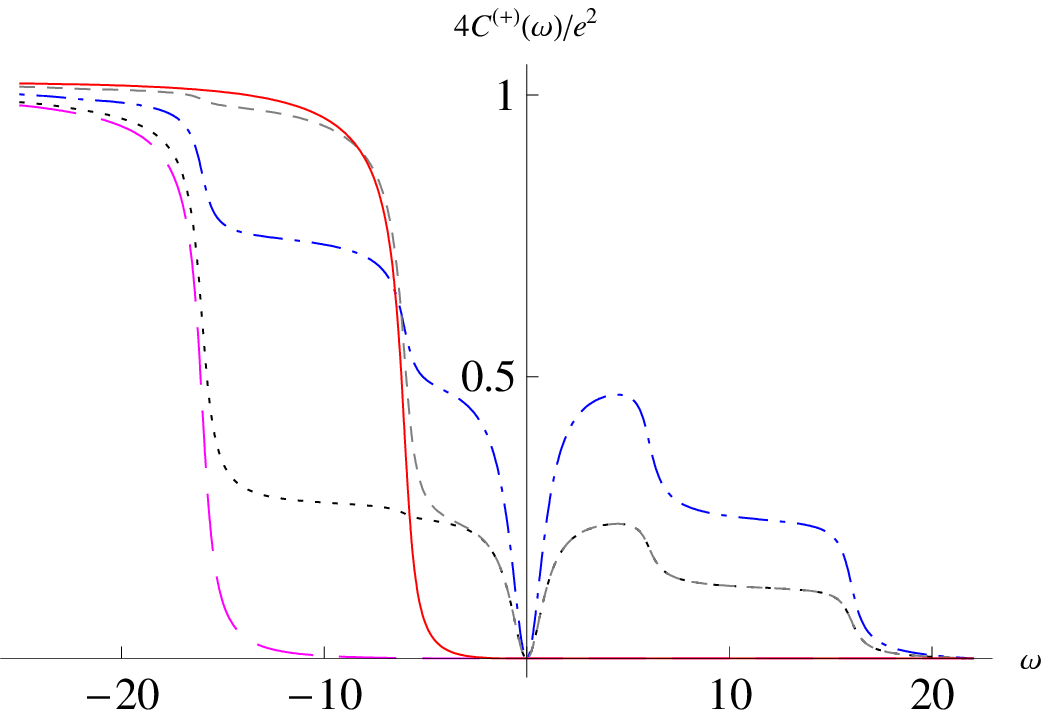}\\
\vspace{0.5cm}\includegraphics[width=2.2in]{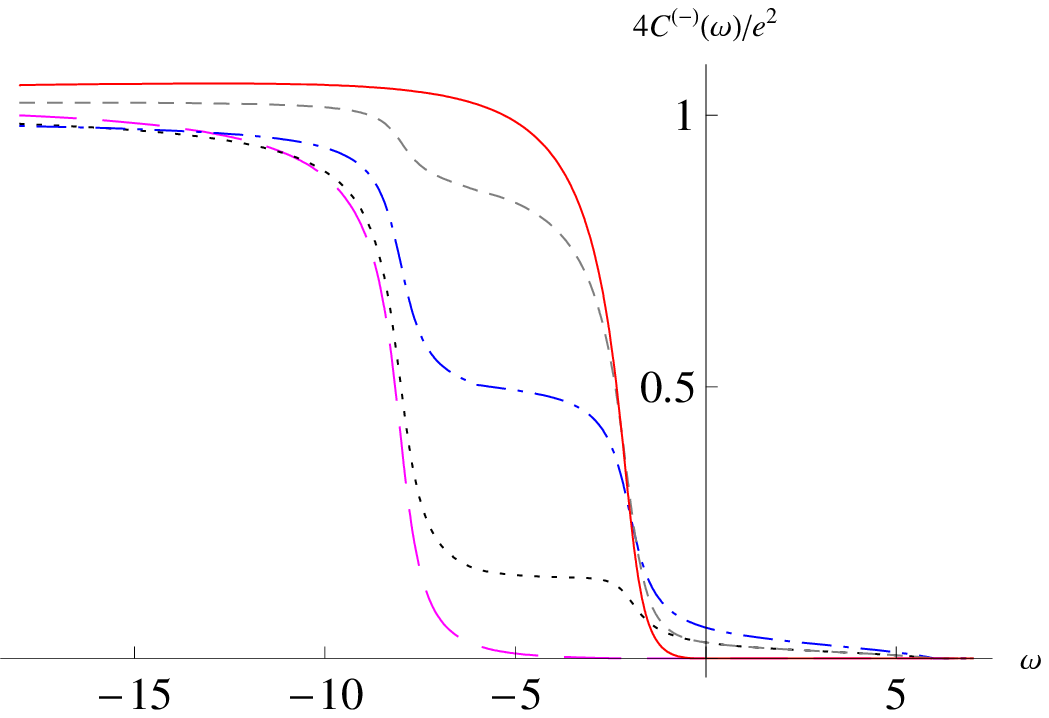}
\hspace{0.5cm}\includegraphics[width=2.2in]{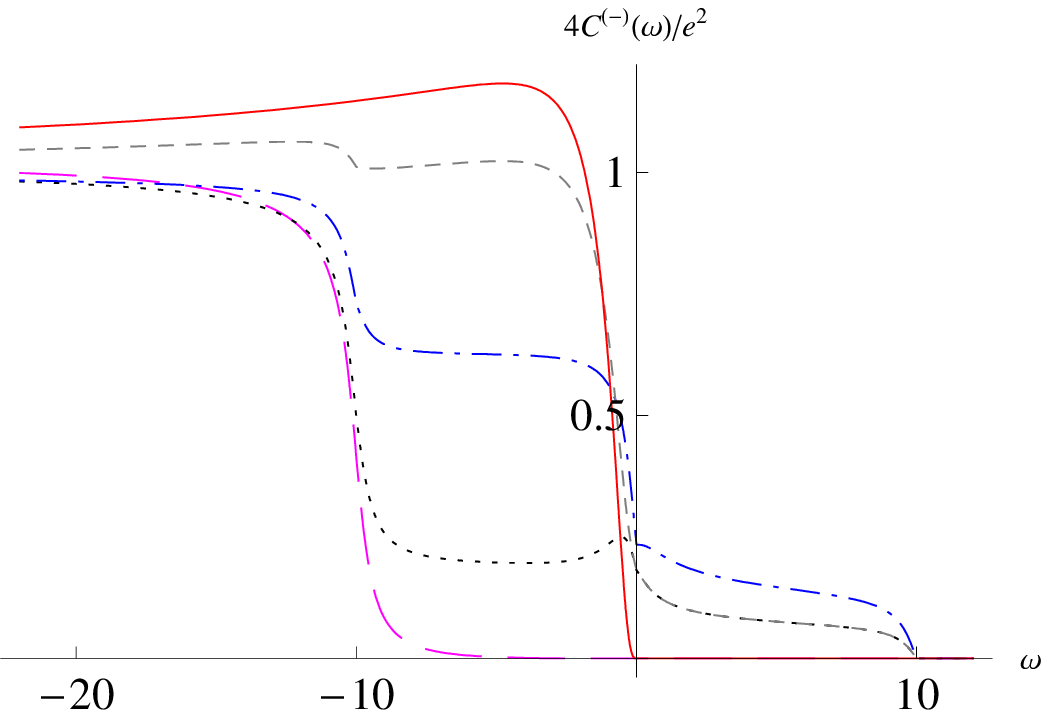}
\hspace{0.5cm}\includegraphics[width=2.2in]{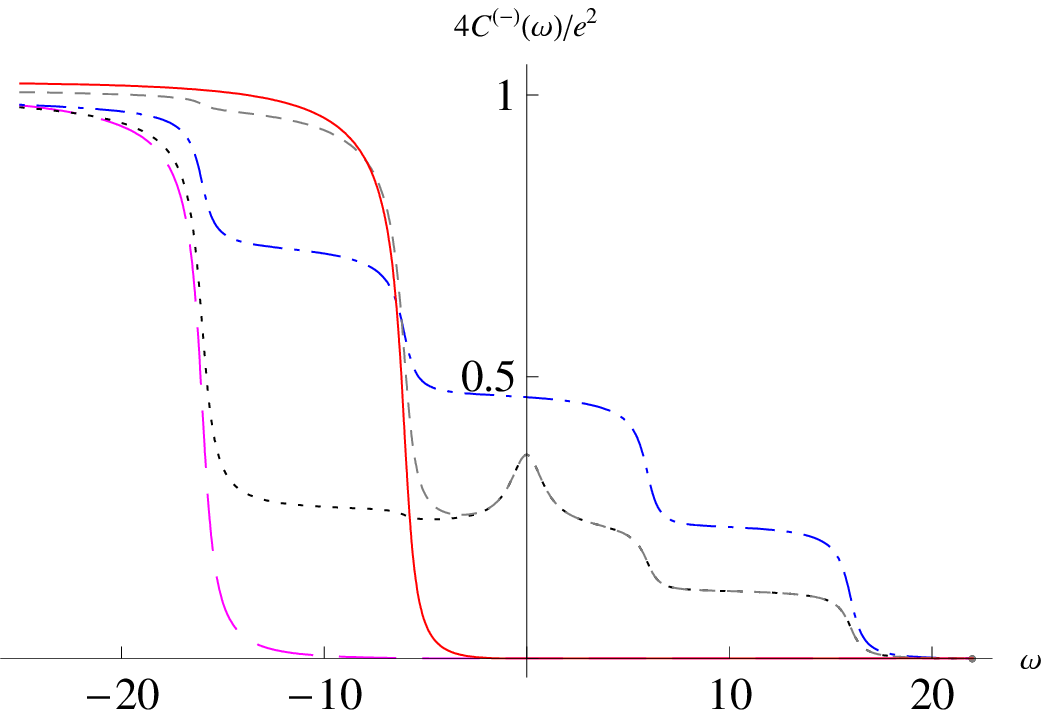}
\caption{(Color online) The noise spectra $C^{(+)}(\omega)$ (upper
panels) and $C^{(-)}(\omega)$ (lower panels) of the biased dot at
zero temperature and $\epsilon^{}_{d}=-5$, for $eV=6,~10$ and $22$
(left, center and right panels). All energies are measured in
units of $\Gamma$. The five curves correspond to zero asymmetry
($a=0$, the dotted-dashed curve), $a=-0.7$ (the dotted curve) ,
$a=0.7$ (the small dashed curve), $a=-1$ (the long dashed curve),
and $a=1$ (the continuous curve).  } \label{NEGEDplus}
\end{figure}


\end{widetext}

When the dot is biased, the contributions of the  four processes
to the noise are all different.  At $T=0$, and using Eq.
(\ref{POT}), Eq. (\ref{NOISE}) becomes
\begin{align}
&\frac{8\pi}{e^{2}}C^{(\pm)}(\omega )=\Theta (-\omega
)\int_{\frac{eV}{2}}^{\frac{eV}{2}-\omega}dE
F^{(\pm)}_{LL}(E,\omega)\nonumber\\
&+\Theta (-\omega
)\int_{-\frac{eV}{2}}^{-\frac{eV}{2}-\omega}dEF^{(\pm)}_{RR}(E,\omega)\nonumber\\
&+\Theta(eV-\omega )\int_{-\frac{eV}{2}}^{\frac{eV}{2}-\omega
}dEF^{(\pm)}_{LR}(E,\omega)\nonumber\\
&+\Theta(-eV-\omega )\int_{\frac{eV}{2}}^{-\frac{eV}{2}-\omega
}dEF^{(\pm)}_{RL}(E,\omega)\ .\label{CVOL}
\end{align}
 Again, we start with $C^{(+)}(\omega)$. From Eq.
(\ref{Fpp}) it follows that each integral in Eq. (\ref{CVOL}) will
generate a step in $C^{(+)}(\omega)$ if a resonance at $E\sim
\epsilon^{}_d$ or at $E\sim \epsilon^{}_d-\omega$ occurs within
the range of integration, and this step will be weighed by the
appropriate product $\Gamma_\gamma\Gamma_{\gamma'}$. The upper
part of Fig. \ref{NEGEDplus} presents results for
$C^{(+)}(\omega)$, for three values of the bias $V$. Comparing
these figures with the curves in Fig. \ref{Cplus} one notes the
following features. (i) As stated above, $C^{(+)}(\omega)$ always
vanishes at $\omega=0$. At small $|\omega|$, the leading
contribution comes from the third term in Eq. (\ref{CVOL}):
\begin{align}
&\frac{4}{e^2}C^{(+)}(\omega)\approx
\frac{8\Gamma_L\Gamma_R}{\pi\Gamma^4}\omega^2\int_{-eV/2}^{eV/2}dE\sin^4\delta(E)\ .\label{eq34}
\end{align}
Using also $\int dE\sin^4\delta(E)=(\Gamma/2)\int
d\delta\sin^2\delta=\delta/2-\frac{1}{4}\sin(2\delta)$, and
assuming that $|\epsilon^{}_d|\ll |eV/2|$ so that the resonance is
fully within the range of integration, we end up again with a
parabolic dip, $4C^{(+)}(\omega)/e^2\approx
(2\Gamma_L\Gamma_R/\Gamma^3)\omega^2$. (ii) Unlike in Fig.
\ref{Cplus}, we now have a finite noise also for $\omega \geq 0$.
This noise arises only from the `$LR$' process, i.e. the third
integral in Eq. (\ref{CVOL}) (note that $\mu^{}_L>\mu^{}_R$). From
Eq. (\ref{Fpp}), the magnitude of this noise is of order
$\Gamma_L\Gamma_R/\Gamma^2=(1-a^2)/4$. It therefore vanishes for
the maximal anisotropy, $|a|=1$, and does not depend on the sign
of $a$ (and therefore the two curves with $a=\pm 0.7$ coincide for
$\omega>0$).  In the two upper left panels, we have
$|\epsilon^{}_d|\ge |eV/2|$, and we observe a plateau in
$C^{(+)}(\omega)$ for $0<\omega<eV$, which decreases back to zero
at both ends of this range.  In contrast, the upper right panel in
Fig. \ref{NEGEDplus} corresponds to $|eV/2| > |\epsilon _d|$. In
that panel we see that for $|a|<1$ the above single plateau splits
into two plateaus, which arise at $\omega=eV/2\pm
|\epsilon^{}_d|$. These steps are due to two resonances which
occur in the `$LR$' process for these frequencies. (iii) For
$\omega<0$, we no longer have symmetry under $a \rightarrow -a$.
For $|a|=1$, we still have a single step in the noise, but this
step now occurs at different frequencies for $a=1$ and $a=-1$. For
$a=1$, this step arises due to the `$RR$' process [second integral
in Eq. (\ref{CVOL})], and it emerges at $\omega =
-|\epsilon^{}_d+eV/2|$. For $a=-1$, this step arises due to the
`$LL$' process [first integral in Eq. (\ref{CVOL})], and it occurs
at $\omega = \epsilon^{}_d-eV/2$. For $|a|<1$, the single step
that appears in Fig. \ref{Cplus} at $\omega\simeq\epsilon^{}_{d}$
splits under the effect of the bias into two steps, located at the
same frequencies as for $a=\pm 1$. The plateau which appears
between these two steps decreases as $a$ decreases from 1 to $-1$.
One may trace this behavior to the different limits of the
integrals, resulting from the different ranges allowed by the
Fermi functions. For example, the `$LL$' process [the first term
in Eq. (\ref{CVOL})] contributes once $\omega$ becomes smaller
than $\epsilon^{}_{d}-(eV/2)$ while the `$RR$' one requires that
$\omega<-|\epsilon^{}_{d}+(eV/2)|$.

We next turn to the current correlations, $C^{(-)}(\omega)$, shown
in the lower part of Fig. \ref{NEGEDplus}. As stated, the
difference $C^{(\times)}(\omega)$ is small for large $|\omega|$.
Indeed, in this range the two rows in Fig. \ref{NEGEDplus} are
very similar. The major differences arise for the shot noise, i.e.
for small
$|\omega|$. As  explained after Eq. (\ref{eq34}), 
the noise for $\omega \geq 0$ is fully due to  the `$LR$' process,
i.e. the third term in Eq. (\ref{CVOL}). This term vanishes at
$|a|=1$, and increases to its maximum as $|a|$ decreases to 0.
When the bias is large enough, $|eV|-|\epsilon^{}_d|>\Gamma$, for
$\omega$ near 0, $C^{(-)}(\omega)$ is determined by the `$LR$'
process, given by Eq. (\ref{integrands}). While the second term is
the second line of Eq. (\ref{integrands})
 is relatively  constant around $\omega=0$, the first term there is significant only when the two resonances overlap,
 i.e. within about $2\Gamma$ of $\omega=0$.
 Since this term is proportional to $a^2$, it
introduces a positive peak in $C^{(-)}(\omega)$ as $|a|$
increases. This peak is indeed clearly seen in the lower right
panel of Fig. \ref{NEGEDplus}.

Unlike $C^{(+)}(\omega)$, which has continuous first and second
derivatives at $\omega=0$, the slope of $C^{(-)}(\omega)$ is {\it
dis}continuous at $\omega=0$ for $T=0$: in this limit the Fermi
distribution can be replaced by a $\Theta$ function [as done in
Eq. (\ref{CVOL})], which has a discontinuous derivative (these
discontinuities also generate discontinuities in the derivatives
at other frequencies, but here we concentrate on the dip or peak
in the shot noise). When $eV=\epsilon^{}_d$, the second integral
in Eq. (\ref{CVOL}), which corresponds to the `$RR$' process, also
exhibits a step at $\omega=0$. This integral, together with its
$\Theta$ function, generate the discontinuity in the derivative of
$C^{(-)}(\omega)$.  Explicitly, we find
\begin{align}
&\frac{8\pi}{e^2}\Bigl[\frac{dC^{(-)}}{d\omega}\Big
|_{\omega\rightarrow 0^+}-\frac{dC^{(-)}}{d\omega}\Big
|_{\omega\rightarrow 0^-}\Bigr
]\nonumber\\&=(36\Gamma_L^2\Gamma_R^2/\Gamma^4)\bigl
(\sin^4\delta(eV/2)+\sin^4\delta(-eV/2)\bigr ).
\end{align}
  Thus, the discontinuities are largest when $|a|=1$ and
when $|\epsilon^{}_d|=|eV/2|$, as can be seen in the lower central
panel in Fig. \ref{NEGEDplus}. This panel also shows a shift of
the peak for $a=-0.7$, to a negative frequency.

\subsection{Finite temperature}

\begin{figure}[h]
\includegraphics[width=2.5in]{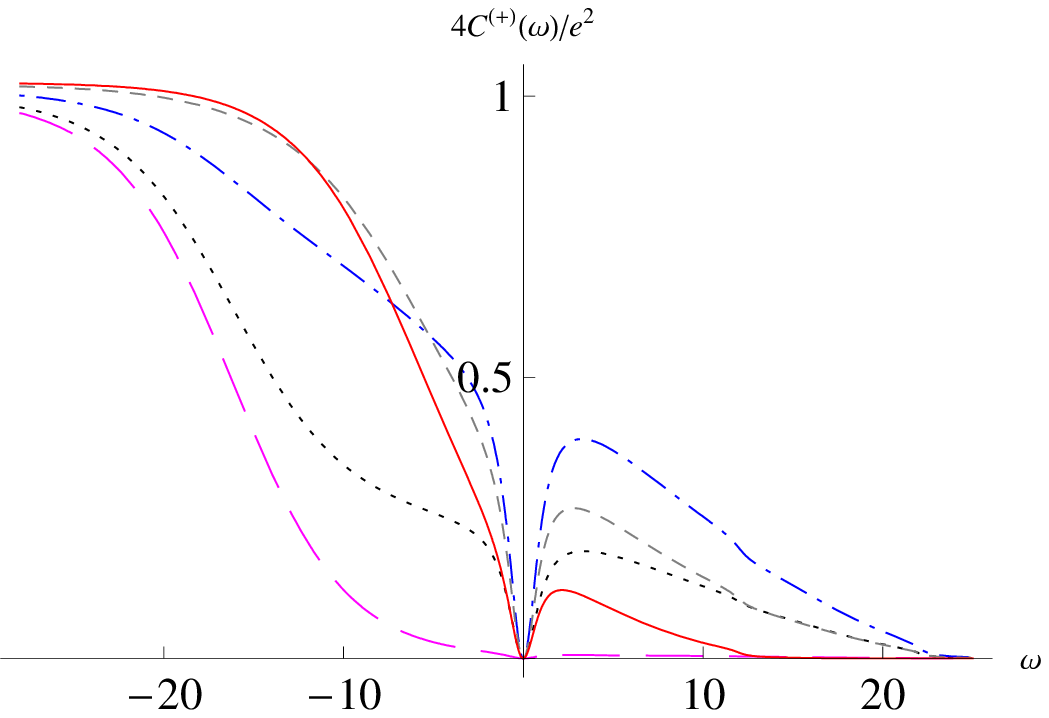}\\
\vspace{0.5cm}\includegraphics[width=2.5in]{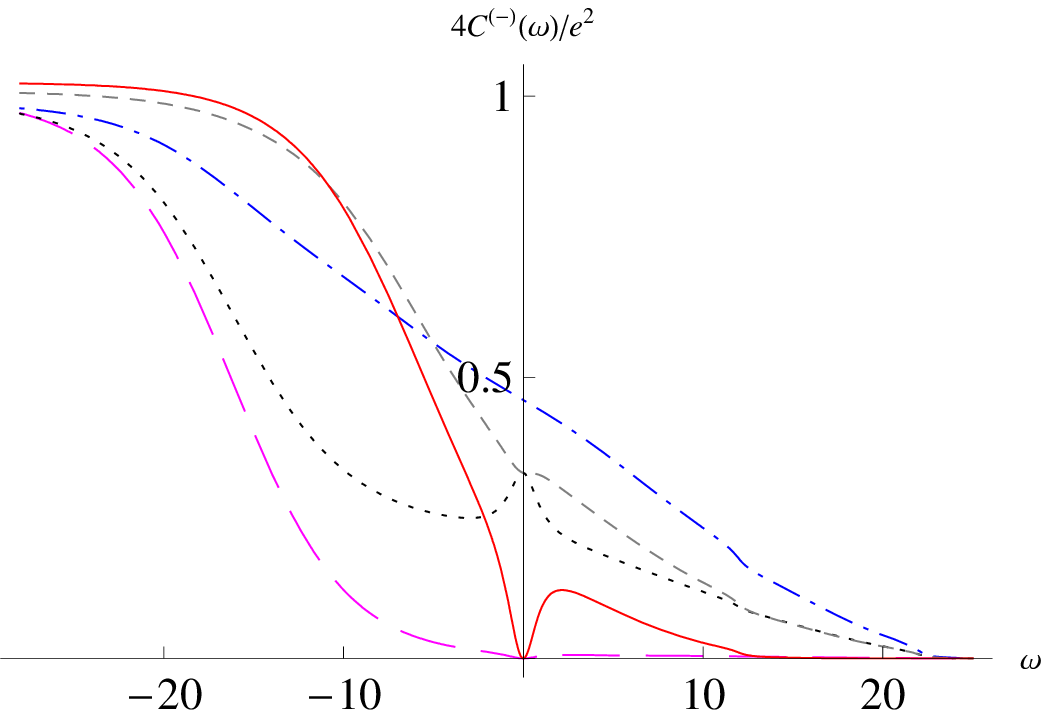}\\
\caption{
(Color online) The two noise spectra of the biased dot at $k_B
T=3$, for $eV=22$ and $\epsilon^{}_{d}=-5$ (all energies and noise
are measured in units of $\Gamma$). The five curves correspond to
zero asymmetry (the dotted-dashed curve), $a=-0.7$ (the dotted
curve) , $a=0.7$ (the small dashed curve), $a=-1$ (the big dashed
curve), and $a=1$ (the continues curve). } \label{FINITET}
\end{figure}

As we showed for $V=0$, finite temperature broadens and smears the
limits of the integrals of Eq. (\ref{NOISE}). Indeed, Fig.
\ref{FINITET} exhibits such a smearing for $eV=22$ and $k_BT=3$.
We show only this value of the bias, since the plots are
qualitatively similar for smaller biases. 
The main new qualitative effect (compared to $T=0$) is the
splitting of the curves for $\pm|a|$ ($a=\pm 0.7$ in Fig.
\ref{FINITET}) at $\omega>0$, due to the contributions of the
`$LL$' and `$RR$' processes there [see Eq. (\ref{Fpp})]. For
$a>0$, the chemical potential $\mu_R$ is closer to $\epsilon^{}_d$
(compared to $\mu_L$). Therefore, the right lead is more strongly
connected to the dot, and this increases the contribution from the
`RR' process, which now integrates over two resonances. This is
responsible for the dip in $C^{(-)}(\omega)$. At high
temperatures, when the `$LL$' process also integrates over two
resonances, the peak in $a=-0.7$ would also turn into a dip.

\section{Summary}

At $T=0$ and $V=0$, the noise spectrum of an unbiased single level
quantum dot exhibits a single step around $\omega\simeq
\epsilon^{}_d$, whose shape depends very weakly on the spatial
asymmetry of the dot or on the fluctuating quantity (current or
charge). In this paper we found how the current and the charge
fluctuations spectra develop additional functional features  when
studied at a finite bias and/or a finite temperature. These
features also depend on the dot asymmetry parameter $a$. Since
this parameter can be varied experimentally, using appropriate
gate voltages, our results suggest several new measurements, which
could yield information on the physics of the quantum dot.

At low temperatures and zero bias, the charge correlation function
$C^{(+)}(\omega)$ should not depend on the asymmetry. However,
even at zero bias, raising the temperature yields a dip around
zero frequency. In contrast, the current correlation function
$C^{(-)}(\omega)$ does depend on the asymmetry, and even at zero
bias and finite $T$ it has a dip around $\omega=0$ whose depth
decreases with decreasing asymmetry.  The details of these dips
may best be observed by measuring the cross-correlations between
the currents on the two leads, $C^{(\times)}(\omega)$.

At finite bias and $T=0$, the single step mentioned above can
split into two or four steps, depending on asymmetry and bias.
Again, the experimental confirmation of the features shown in Fig.
\ref{NEGEDplus} can also give information on the location of the
resonance and on its partial widths. It would be particularly
interesting to study the various noise functions in the shot-noise
region, where we find a variety of dips and peaks. At finite
temperatures we also predict that for a finite asymmetry some dips
can turn into peaks.

\vspace{.5cm}

\begin{center}{\bf Acknowledgement}\end{center}

 We thank Y. Imry, S. Gurvitz, A. Schiller,
D. Goberman, V. Kashcheyevs, and V. Puller for valuable
discussions. This work was supported by the German Federal
Ministry of Education and Research (BMBF) within the framework of
the German-Israeli project cooperation (DIP), and by the Israel
Science Foundation (ISF).

\end{document}